\documentclass[final,3p,times,twocolumn]{elsarticle}
\biboptions{comma,sort&compress}

\usepackage{graphicx}
\usepackage{epsfig}
\usepackage[utf8x]{inputenc}
\usepackage{pstricks}
\usepackage{xspace}
\usepackage{subfigure}

\usepackage{amsmath}
\usepackage{amssymb}
\def\nin{\noindent}
\def\beq{\begin{equation}}
\def\eeq{\end{equation}}
\def\bea{\begin{eqnarray}}
\def\eea{\end{eqnarray}}

\def\alar{$\bar{\Lambda}/\Lambda$\xspace}

\def\ks{$\mr{K^0_S}$\xspace}
\def\pt{$p_{\mr{T}}$\xspace}
\newcommand{\mr}{\mathrm}
\journal{Nuc. Phys. (Proc. Suppl.)}

\begin{document}

\begin{frontmatter}



\title{First minimum bias physics results at LHCb}

 \author{Francesco Dettori\corref{cor1}}
 \address{Università degli Studi di Cagliari and INFN Sezione di Cagliari\\ 
Dipartimento di  Fisica, Cittadella Universitaria
 09042 Monserrato (CA) Italy
}
\cortext[cor1]{On behalf of the LHCb Collaboration}
\ead{francesco.dettori@cern.ch}

%\author{}

%\address{}

\begin{abstract}
\noindent

We report on the first measurements of the LHCb experiment, as obtained 
from $pp$ collisions at $\sqrt{s} = 0.9$~TeV and 7 TeV recorded using a minimum bias trigger. 
In particular measurements of the absolute \ks production cross section at $\sqrt{s} = 0.9$~TeV  and of the \alar ratio both at $\sqrt{s} = 0.9$~TeV and 7 TeV are discussed  and preliminary results are presented.

\end{abstract}


\end{frontmatter}

%\linenumbers

\section{Introduction\label{sec:intro}}

The production of strange quarks offers a very good testing ground
for phenomenological models of proton-proton interactions~\cite{Skands:2010ak}. 
In particular, being the strange quark mass at the level of $\Lambda_{QCD}$ it probes the non perturbative QCD regime.
Moreover, as the valence quarks in the proton are \emph{up} and \emph{down} only, the study of strangeness production provides
 a good environment to assess and tune phenomenological models of hadronisation, thereby improving Monte Carlo generators at LHC energies.  

For 2009 and the first few months of operation in 2010 the LHC ran at very low luminosity.  This allowed LHCb to deploy an
  efficient and simple minimum bias trigger. Data sets collected in 2009 and early 2010, at $\sqrt{s}=0.9$ TeV and 7 TeV, allow then the study of the minimum bias sector. 
Moreover LHCb profits from its unique rapidity coverage in the forward region, with pseudorapidity acceptance of approximately  $2 < \eta < 5$, and detection capability to very low transverse momentum. 

The absolute production of \ks in $pp$ collisions and the \alar ratio as measured with the LHCb detector at LHC 
is presented in this paper, showing preliminary results. The former has been measured on 2009 data taken at $\sqrt{s}=0.9$~TeV while the latter has been studied on 2010 data both at $\sqrt{s}=0.9$~TeV and 7~TeV.

\section{The LHCb experiment and detector \label{sec:detector}}
\nin
The LHCb experiment at LHC will study CP violation and rare decays, and CKM matrix tests will be performed on the full $b$-hadrons spectrum as well as in
the $charm$ sector in search for hints of new physics \cite{LHCb:2008,LHCbroadmap}.

The LHCb detector is a single-arm spectrometer placed in the forward region of
the $pp$ interaction point, with an angular coverage, with respect to the
beam-axis, from 10 (15) to 300 (250) mrad in the bending (non-bending) plane.
% suited to collect the $b \bar b$
% quarks production which, at the LHC energies, is well peaked and correlated in
% the forward and backward region. 
Here, just a brief description of the detector is given, more details can be found in Ref.~\cite{LHCb:2008}.

The LHCb detector is divided in several subdetectors. Around the interaction point is the VELO, a silicon detector that measures the radial and azimuthal hit points from charged tracks. The momentum measurement is ensured by the dipole magnet and the tracking system. The latter is subdivided in a two silicon  
micro-strips detectors stations (TT) placed upstream of the magnet, and three stations (T1-T3), downstream of the magnet, made up of silicon micro-strips for the inner part and straw tubes for the outer part.
Particle identification is provided by two Ring Imaging Cherenckov detectors, RICH1 and RICH2, while the MUON system is composed by one detector station (M1) placed upstream of the calorimeter system and 4 downstream (M2-M5); the stations are
build up from MWPC's with the exception of the inner part of M1 where
triple-GEM detectors are exploited. Finally, energy measurement is made by the calorimeter system: a Scintillator Pad Detector (SPD) and Pre-Shower (PS), and the Electromagnetic (ECAL) and Hadronic (HCAL) calorimeters.

\section{Data taking conditions}
\nin
During the 2009 run, low-intensity proton beams were collided in LHCb at $\sqrt{s} = 0.9$ TeV. In these conditions both the crossing angle and size of the beams are larger than the nominal, at higher energies, for which the detector has been designed. For these reasons, in order not to risk damage in the VELO, its two halves were kept 15 mm away from the nominal closed position, reducing the azimuthal acceptance of the detector. For 2009 run the magnetic dipole field was in the down position. 

Data in 2010 were taken both at $\sqrt{s} = 0.9$ TeV and 7 TeV and at both field polarities. 
At  $\sqrt{s} = 0.9$ TeV the same reasons as for 2009 meant that data had to be taken with the VELO in a semi-open position, but this time the two halves were kept 10 mm away from their nominal closed position; at 7 TeV data were taken with VELO in the fully closed position. 

A very loose minimum bias trigger was set up in order to trigger inelastic $pp$ collisions. 
In 2009 this trigger was provided by the calorimeter system, while in 2010 a requirement was made of at least one charged track in the VELO detector or downstream tracking system.
Moreover, also beam-gas interactions for both LHC proton beams were recorded; these data are crucial for the measurement of the beam-size needed to calculate the luminosity. 

\section{\ks production cross-section}
\begin{figure}
\includegraphics[width  = \columnwidth, height = 5 cm]{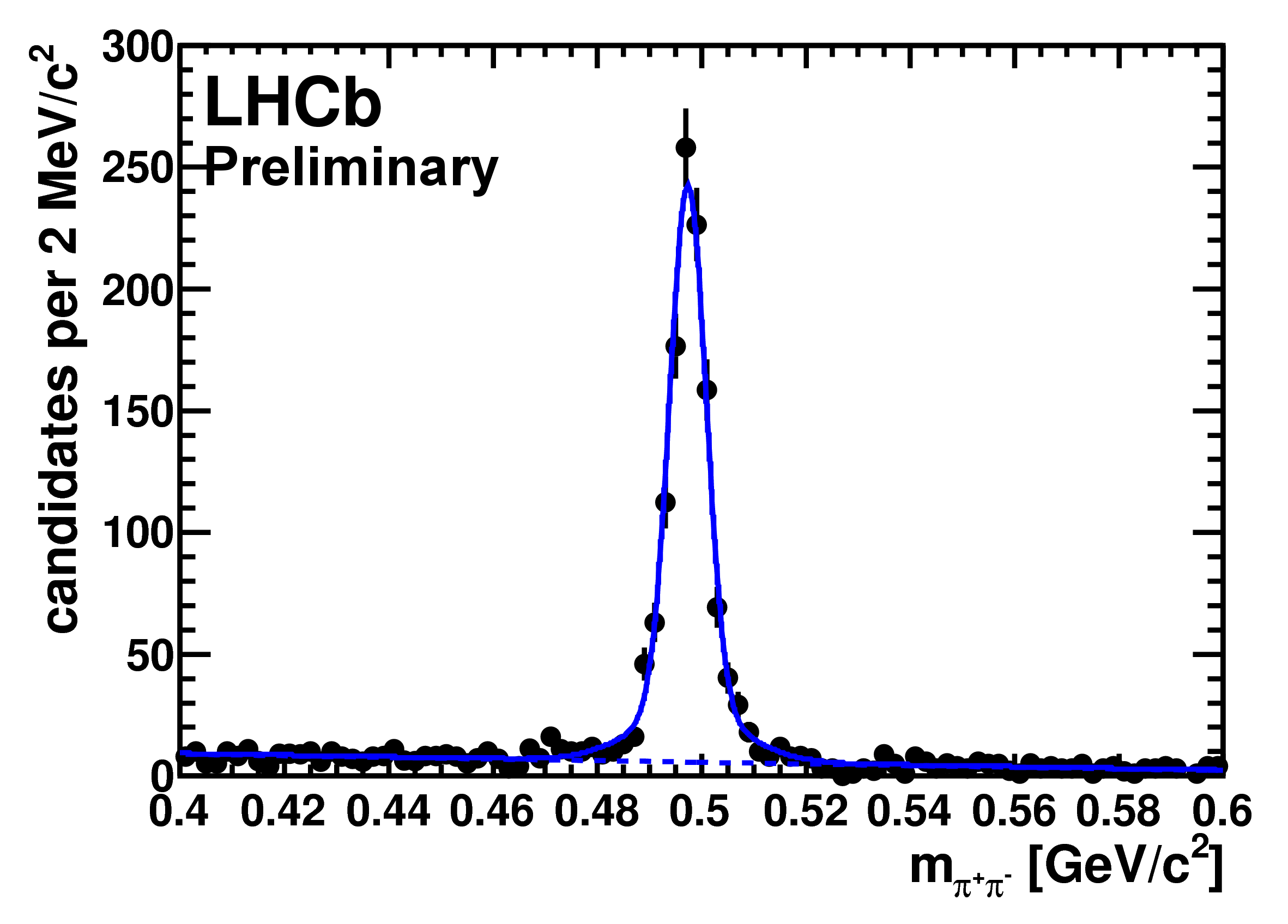}
\caption{Invariant mass distribution for $\mr{K^0_S} \to \pi^+ \pi^-$ candidates as reconstructed from \emph{long} tracks after selection. Beam-gas interactions have been statistically subtracted from this distribution. }\label{fig:ks_masses}
\end{figure}

\nin
\ks candidates were reconstructed from the $\pi^+ \pi^-$ decay mode.
Due to the non standard position of the VELO and long \ks lifetime, only a fraction of the decays could leave a signal in the VELO. Therefore, two parallel analyses were carried out. The first was based on tracks that left hits in the entire tracking system (VELO, TT and T1-T3), called \emph{long} tracks, while the second was based on \emph{downstream} tracks, which do not require any information from the VELO. 
% The analysis based on \emph{downstream} tracks required the \ks candidate to point to the $z$ (beam) axis, while, long tracks \ks were required to point to the reconstructed PV. In the long tracks selection the main separation variable to discriminate combinatorial background is defined as:
% \beq
% \nu= \ln \left(\frac{IP_1\cdot IP_2}{IP_{\mr{K^0_S}}\cdot (1~\mr{mm})} \right)
% \label{eq:nu}
% \eeq 
% where $IP_{1,2}$ and $IP_{\mr{K^0_S}}$ are the impact parameters with respect to the PV of the daughter particles and of the \ks respectively. 
% Further constraints were required for both analyses on the $\chi^2$ of the tracks fits and on the $\chi^2$ of the \ks decay vertex. 
The invariant mass distribution for \ks candidates after selection is shown in Fig.~\ref{fig:ks_masses} for the \emph{long} tracks analysis.
\ks production yields have been measured in bins of rapidity ($y$) and transverse momentum (\pt). 
For each bin the signal estimation has been made by means of a $\chi^2$ fit using a linear function to parametrise the background and a double gaussian with common mean to represent the signal. 
In order to estimate the reconstruction and selection efficiency Monte Carlo (MC) simulated $pp$ collisions have been generated with the PYTHIA 6.4~\cite{Sjostrand:2006za} package and propagated through the detector with full simulation with the GEANT 4 package~\cite{Agostinelli:2002hh}. 
The trigger efficiency has been estimated from the MC simulations, and cross-checked directly from data, exploiting events triggered independently from \ks daughters particles. 
The tracking efficiency was calculated from MC detailed detector simulations, which have been corrected in order to reproduce data hit-finding inefficiency due to residual misalignment. 
%The systematic uncertainty associated with this correction has been estimated measuring the tracking efficiency directly on the data by splitting the tracks in different segments. 
The selection efficiency has been estimated both in data and MC and compared using samples selected with a pre-selection looser than the final one. 
Finally, systematic uncertainties for non-prompt \ks and diffractive events modelling in MC simulations have been estimated and associated to the final results.

\subsection{Luminosity measurement}
\nin
The luminosity has been measured using a direct method \cite{FerroLuzzi:2005em}. It can be expressed as 
\beq
\mathcal{L} = n_1 n_2 f 2 c \cos^2 \theta \int \rho_1 \rho_2 \mr{d}V \mr{d}t
\eeq
where $n_{1,2}$ are the number of protons per bunch, $\theta$ is the beam-crossing half angle and the integral computes the overlap of the beams during a bunch crossing. 
%Since the bunch shapes are well described as three-dimensional gaussians, only sections and mean positions had to be measured. 
Vertices from beam-gas interactions have been used to measure the transverse beam sections and crossing angle, while the luminous region parameters (where $pp$ collisions occur) have been measured from $pp$ primary vertex (PV) positions. 
Finally, the bunch intensities were determined using beam current transformers (BCT), under the responsibility of the accelerator beam instrumentation division.
This measurement dominates the final uncertainty on the luminosity, contributing with a relative error of 12\%. 
The total integrated luminosity for the entire sample considered in the analysis was found to be $\mathcal{L}_{\mr{int}} = 6.8 \pm 1.0 ~\mu \mr{b}^{-1}$.

\subsection{Results}
\nin
The cross-sections for \ks production were evaluated for both the \emph{downstream} and \emph{long} analysis. The results of the two are in agreement with each other. As the two sets of measurements have a significant correlation, it has been decided to quote the most precise result in each bin. In general this means that the result of the downstream analysis is chosen, but for some bins at low $p_T$ the long analysis has better acceptance and hence is used for the final result.

\begin{figure}[!h]
\includegraphics[width  = \columnwidth]{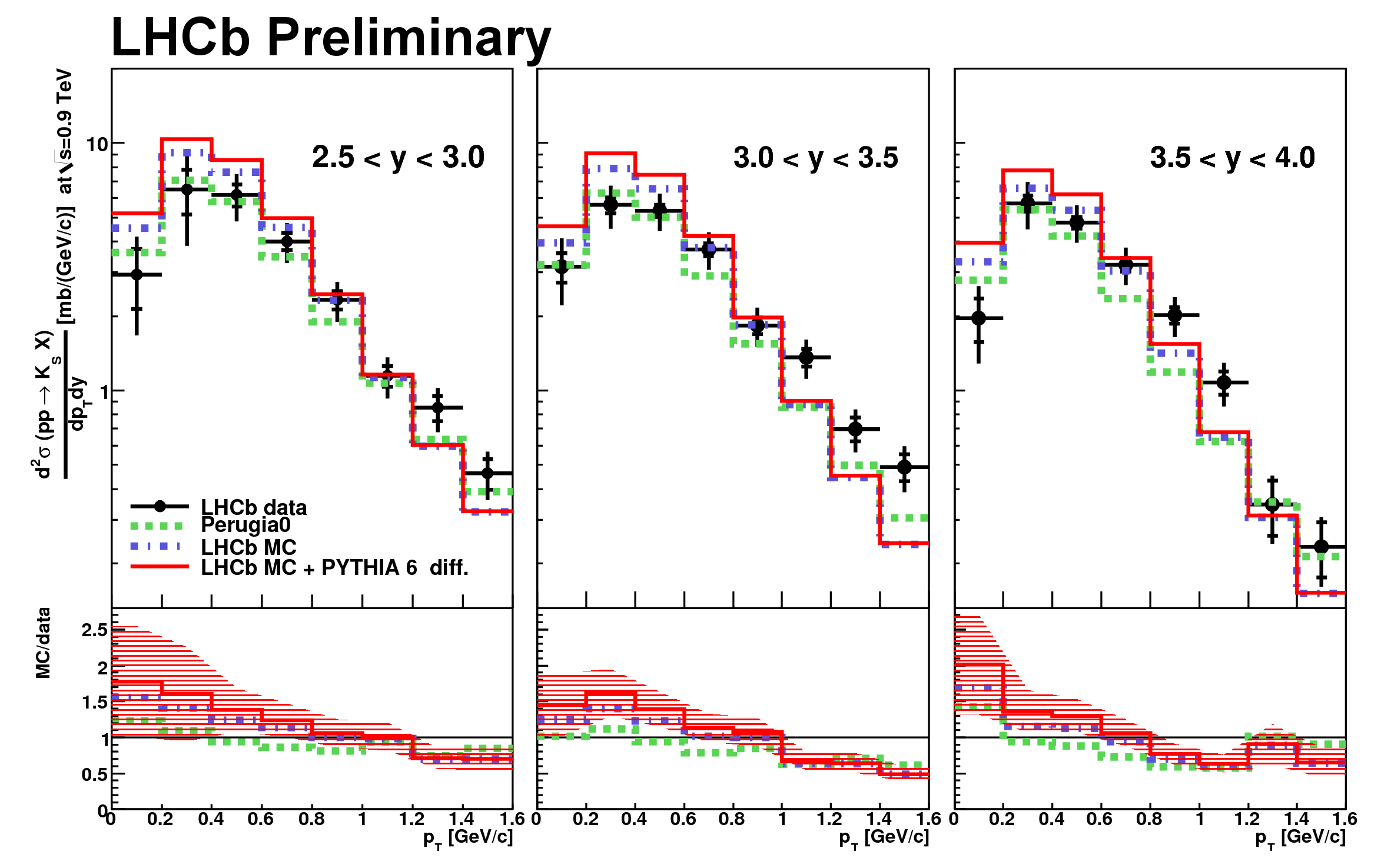}
\caption{Double differential prompt \ks production cross-section in $pp$ collisions at $\sqrt{s}=0.9$ TeV. 
Data are presented as three transverse momentum distributions in three bins of rapidity.
LHCb preliminary results are marked by black dots: error bars represent statistical uncertainties while thick bars represent the systematic ones. Data are compared to predictions from different settings of the PYTHIA generator \cite{Sjostrand:2006za, Skands:2010ak}. Bottom plots show MC to data ratios.  }
\label{fig:ks_results}
\end{figure}

\begin{figure}
\includegraphics[width  = \columnwidth, height = 5 cm]{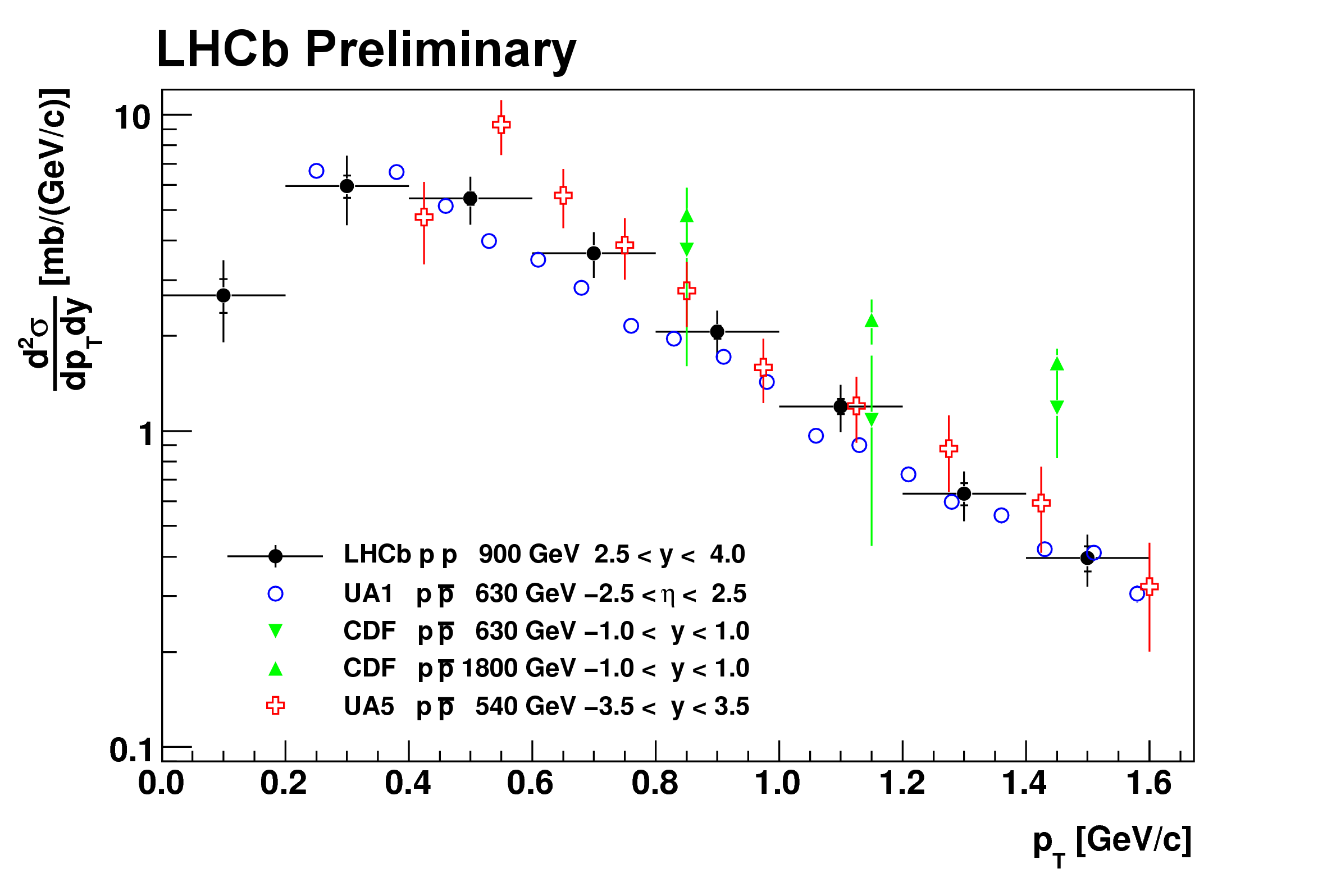}
\caption{Comparison of double differential \ks production cross-section measurements obtained by 
the LHCb, CDF~\cite{Abe:1989hy}, UA1~\cite{Bocquet1996441}, UA5~\cite{Alner:1985ra} experiments in different rapidity ($y$) or pseudorapidity ($\eta$) ranges.}\label{fig:ks_res_comp}
\end{figure}

In Fig.~\ref{fig:ks_results} the results are shown as three transverse momentum spectra for the different rapidity ranges. Data is compared to LHCb Monte Carlo and  PYTHIA 6 MC generator with the so-called ``Perugia 0'' settings \cite{Sjostrand:2006za, Skands:2010ak}. The agreement with MC is reasonable, although the data appear to have a slightly harder \pt spectrum. 
In Fig.~\ref{fig:ks_res_comp}  the LHCb data are compared to different measurements made by the
CDF~\cite{Abe:1989hy}, UA1~\cite{Bocquet1996441}, UA5~\cite{Alner:1985ra} experiments in different rapidity ($y$) or pseudorapidity ($\eta$) ranges. The agreement is good, and it can be seen that the LHCb results are valuable in extending the measurement set both to lower $p_T$ and higher values of rapidity.

%\cite{Alner:1985ra, Drijard:1982, FerroLuzzi:2005em}

\section{\alar production ratio}
\nin
The \alar ratio was studied in data recorded in 2010 by the LHCb detector. Data collected at  $\sqrt{s} = 0.9$ TeV correspond approximately to an integrated luminosity $\mathcal{L_{\mr{int}}} \simeq 0.31 ~\mr{nb^{-1}}$, while those collected at $\sqrt{s}=7$ TeV to an integrated luminosity of $\mathcal{L_{\mr{int}}} \simeq 0.2 ~\mr{nb^{-1}}$.  

\begin{figure}[!h]
 \centering \includegraphics[width=\columnwidth, height = 5 cm]{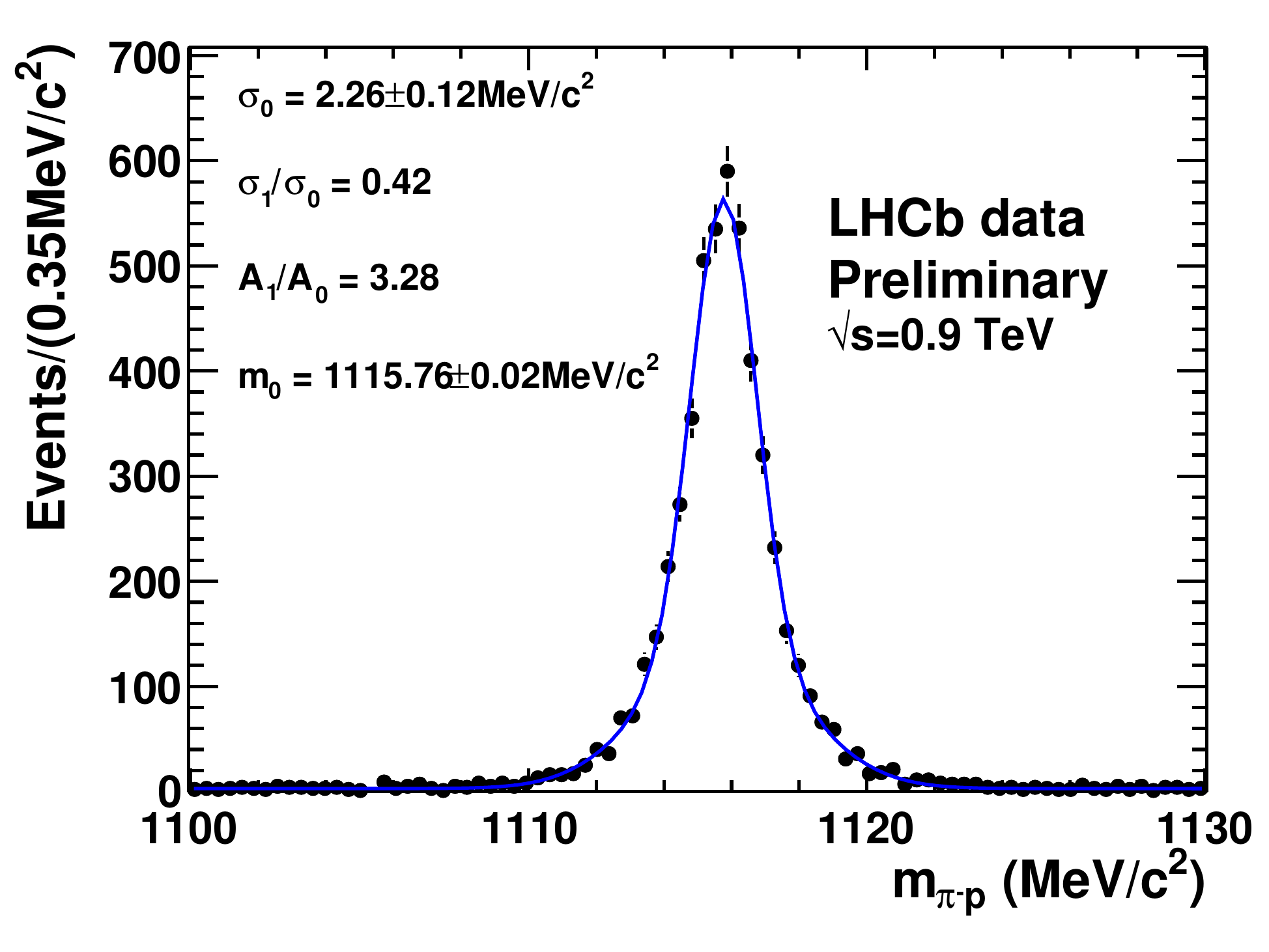}
 \caption{Invariant mass distribution for $\Lambda \to p \pi^-$ candidates after selection, at $\sqrt{s}=0.9$ TeV. }
  \label{fig:lam_mass}
\end{figure}

$\Lambda$ particles and antiparticles were reconstructed in the $\Lambda (\bar \Lambda) \to p^{+(-)} \pi^{-(+)}$ decay modes.
Pairs of \emph{long} tracks were selected requiring the track fit $\chi^2/\mr{ndof}$ to be less than 20; they were then combined together to make a $\Lambda$ candidate which was accepted if constraints on vertex quality and on pointing angle of the mother
to the PV were satisfied. 
In addition a discriminat variable based on the impact parameters of the track candidates with respect to the PV, $IP_1$ and $IP_2$, and of the $\Lambda$ candidate itself, $IP_{\Lambda}$ was defined: 
\begin{equation}
\nu = \log \left(\frac {IP_1 \cdot IP_2}{IP_{\Lambda} \cdot (1 \mr{mm})}\right) 
\end{equation}
and used to reject combinatoric background. No particle identification was used, therefore, in order to remove the background coming from \ks reflections, the \ks invariant mass region in the $\pi^+ \pi^-$ hypothesis was cut away. The $\Lambda$ invariant mass distribution after selection is shown in Fig.~\ref{fig:lam_mass}. 
The data samples divide approximately equally into \emph{field-up} and \emph{field-down}, in which the dipole field is pointing in opposite directions; the results for \alar ratio were found to be consistent with each other,
indicating no significant charge-dependent systematics,  and averaged for the final results. 

Different sources of systematic uncertainties were considered: MC modelling of diffractive and non-prompt contributions; $\Lambda$ production and absorption along the flight path; variation of selection cuts; influence of $\Lambda$ transverse polarisation on acceptance; $p$ and $\bar p$ interactions in the detector;  $p_{\mr{T}}$ calibration of the efficiency. Since the final states of $\Lambda$ and $\bar \Lambda$ are kinematically equivalent many effects cancelled out. The sum of the systematic uncertainties amounts to 2\% relative plus a 0.02 absolute uncertainty.

\begin{figure}[!h]
\begin{center}
%\subfigure[\alar results at $\sqrt{s}=0.9$ TeV]{
\includegraphics[width  = \columnwidth, , height = 5 cm]{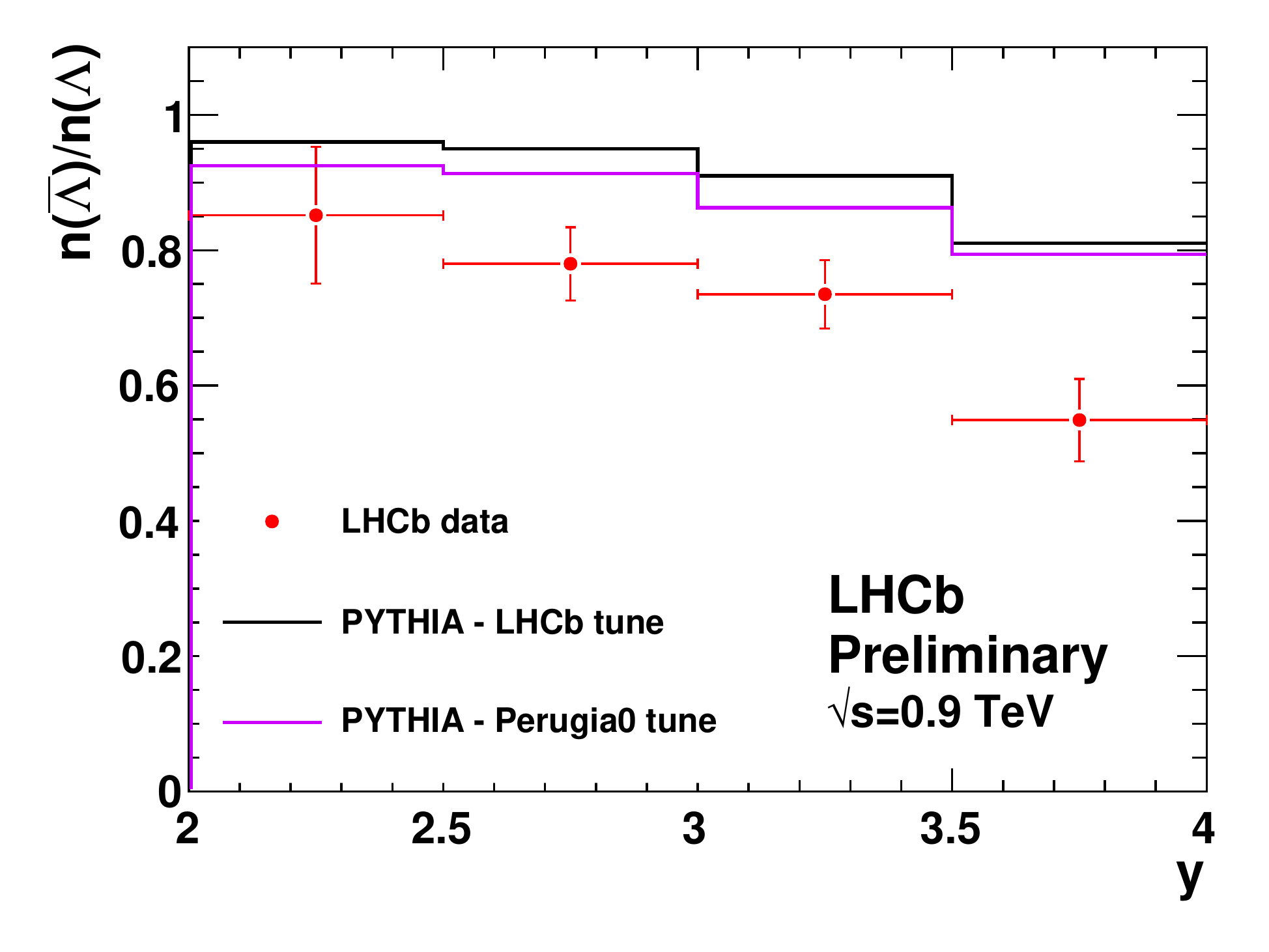}
%}\subfigure[\alar results at $\sqrt{s}=7$ TeV]{
\includegraphics[width  = \columnwidth, , height = 5 cm]{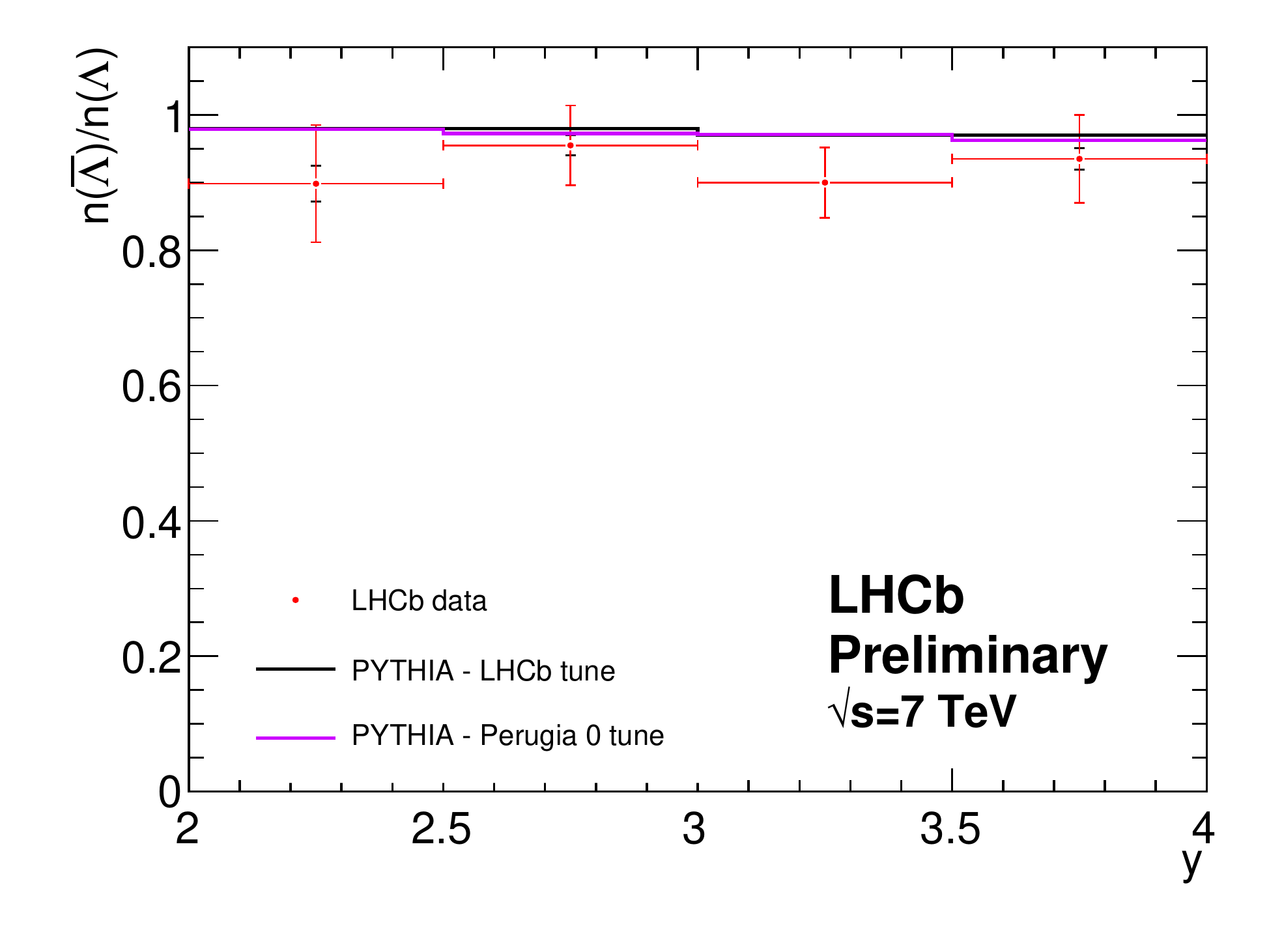}%}
\end{center}
 \caption{Summary of  results on \alar at $\sqrt{s}=0.9$ TeV (top) and  at $\sqrt{s}=7$ TeV
 (bottom). For the data points the error bars are the quadratic sum of
 statistical and systematic uncertainties. Data is compared to LHCb Monte Carlo and  PYTHIA 6 MC generator with the so-called ``Perugia 0'' settings \cite{Skands:2010ak, Sjostrand:2006za}.}
  \label{fig:lam_results}
\end{figure}

\begin{figure}[!h]
 \includegraphics[width  = \columnwidth, height = 5 cm]{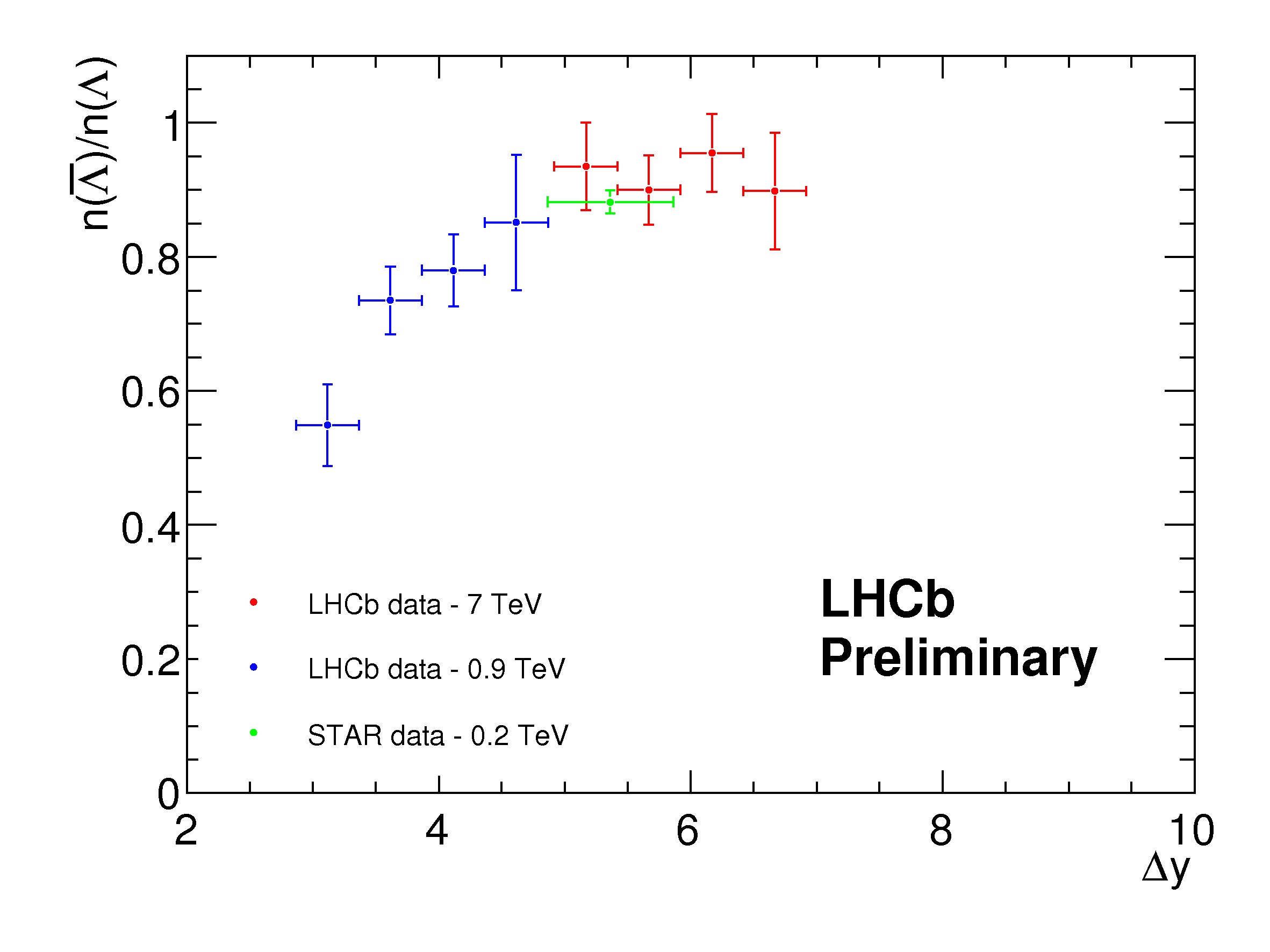}
\caption{Summary of results on \alar at $\sqrt{s}=0.9$ TeV and  at $\sqrt{s}=7$ TeV versus $\Delta y = y_{\mr{beam}} - y_{\Lambda}$. LHCb data is compared to one data point measured by the STAR experiment \cite{PhysRevC.75.064901}. }\label{fig:deltay}
\end{figure}

The final results are shown in Fig.~\ref{fig:lam_results}. The data are compared with the LHCb MC and PYTHIA 6 MC generator simulations with the so-called ``Perugia 0'' settings \cite{Skands:2010ak,Sjostrand:2006za}. It can be seen that the measured ratio at $\sqrt{s}=0.9$~TeV is significantly below the prediction of the generators.
The same results are also shown in Fig.~\ref{fig:deltay}, as a function of $\Delta y = y_{\mr{beam}} - y_{\Lambda}$, which allows the comparison of results from different rapidity regions and different beam energies. 
The LHCb data are seen to be consistent with a continuous distribution over a wide range in $\Delta y$. They are also consistent with a previous measurement provided by the STAR experiment. \cite{PhysRevC.75.064901}.

%\vfill\eject
\bibliographystyle{utphys}
\bibliography{Bibliography}

\end{document}